\documentclass[12pt,draftcls,onecolumn]{IEEEtran}

\ifCLASSOPTIONcompsoc
\usepackage[nocompress]{cite}
\else
\usepackage{cite}
\fi

\usepackage{amsmath,amsthm,amssymb}
\usepackage{array}
\usepackage{graphicx}
\usepackage{subfig}
\usepackage{fancyhdr}
\usepackage{color}
\usepackage{extarrows}
\usepackage{enumerate}
\usepackage{epstopdf}
\newtheorem{definition}{Definition}
\newtheorem{theorem}{Theorem}

\newtheorem{remark}{Remark}

\begin{document}
\title{Lyapunov Functions in Piecewise Linear Systems: From Fixed Point to Limit Cycle}

\author{Yian Ma, Ruoshi Yuan, Yang Li, Ping Ao, Bo Yuan
\thanks{This work was supported in part by the National 973 Projects No.~2010CB529200 and by the Natural Science Foundation of China No.~NFSC61073087 and No.~NFSC91029738.}
\thanks{Y. Ma, R. Yuan, Y. Li were with the Department of Computer Science and Engineering, Shanghai Jiao Tong University, when this work is performed.
Currently, Y. Ma is with the Department of Applied Mathematics, University of Washington.
R. Yuan is with the Shanghai Center for Systems Biomedicine,
Shanghai Jiao Tong University.
L. Yang is with the Department of Statistics, Harvard University.}
\thanks{P. Ao is with Shanghai Center for Systems Biomedicine and Department of Physics, Shanghai Jiao Tong University (e-mail: aoping@sjtu.edu.cn).}
\thanks{B. Yuan is with Department of Computer Science and Engineering, Shanghai Jiao Tong University (e-mail: boyuan@sjtu.edu.cn).}
}


\maketitle

\begin{abstract}

This paper provides a first example of constructing Lyapunov functions in a class of piecewise linear systems with limit cycles.
The method of construction helps analyze and control complex oscillating systems through novel geometric means.
Special attention is stressed upon a problem not formerly solved:
to impose consistent boundary conditions on the Lyapunov function in each linear region.
By successfully solving the problem, the authors construct continuous Lyapunov functions in the whole state space.
It is further demonstrated that the Lyapunov functions constructed explain for the different bifurcations leading to the emergence of limit cycle oscillation.
\end{abstract}


\IEEEPARstart{P}{iecewise} linear systems (PLS), as a kind of hybrid systems, have attracted wide interest.
Much study has been done on this class of systems to understand the complex behaviors of the nonlinear systems and make control possible.
On one hand, a wide variety of natural and technological systems are frequently modeled in PLS, such as neural oscillators \cite{Neural1,Neural2}, hopping robots \cite{Hop}, and control systems subject to actuator saturation \cite{Hu02saturate}.
On the other hand, PLS introduce nonlinearity more easily and more controllably for performance improvement and can act as the dynamical inclusion of other nonlinear systems \cite{Hu04stabilityregions}.
Hence, stability and performance issues are hotly discussed in the literature.

As the ultimate criteria for stability and robustness analysis \cite{Sastry}, optimal control \cite{Zinober}, and system identification, Lyapunov function is often constructed to analyze the dynamic properties of the PLS.
There has been much study on its construction in the complete phase space \cite{Rantzer2000}.
Previous works systematically reshape the Lyapunov function of each linear region in order to have Lyapunov function coincide at the boundaries.
However, when limit cycle oscillation emerges, the function constructed in different regions
is not continuous in the whole phase space, and hence, fails to posses the property of a Lyapunov function.

The objective of this paper is to provide a first example of constructing Lyapunov functions in a class of PLS with limit cycle oscillation.
The novel approach tackles a central obstacle faced by previous efforts: Lyapunov function of different linear regions does not equal to each other on the boundaries.
The Lyapunov function constructed using the new methodology accounts for: asymptotic stability of the fixed points, stable regions of the whole system, and the process through which transient states settle into stable oscillation.

Moreover, the Lyapunov functions constructed for the class of PLS offer a geometric view of the feedback control systems.
The change of the geometric configuration of the Lyapunov function figuratively describes the evolution of systems' dynamics.
For the class of systems discussed in this paper, change in the Lyapunov function explains for the two different bifurcation phenomena of a system into oscillation: Hopf bifurcation and SNIP (Saddle-Node-Infinite-Period Bifurcation) bifurcation.

\section{Previous Works}
Previous works generally consider analysis of PLS of the form \cite{Rantzer}:
\begin{align}
\dot{\mathbf{x}}(t)=\mathbf{f}_i(\mathbf{x})=A_i\mathbf{x}(t)+\mathbf{a}_i
\end{align}
for $\mathbf{x}(t)\in M_i$.
Here, $M_i\subseteq\mathbb{R}_i$ is a partition of the state space into a number of polyhedral cells.

For the record, the first generic approach in constructing a Lyapunov function for a piecewise linear model is the piecewise quadratic Lyapunov function (QLF) \cite{Rantzer}.
This piecewise QLF method constructs a quadratic Lyapunov function in each linear region, and fit each of the pieces together on the boundaries.
This approach is uniform and computationally tractable when the systems contain only fixed points as limit sets.
However, when the systems have oscillating behavior, piecewise QLF can no longer be continuous over the boundaries \cite{SLF}.
Since complex behaviors like oscillation are quite prevalent in natural and technological systems \cite{winfree,LimitCycleControl}, a method of constructing Lyapunov functions is needed for the analysis and control of them.

To apply the Lyapunov function criteria to the complex situations, many efforts \cite{Review} have attempted to modify the piecewise QLF approach, such as finding multiple QLFs \cite{Branicky1998}, obtaining a QLF outside the LaSalle invariant set \cite{Hu04stabilityregions}, and constructing a surface QLF \cite{SLF}.
But none have yet addressed the boundary issue to make Lyapunov function continuous in the whole phase space.

\begin{figure}
\begin{center}
  \includegraphics[width=0.9\textwidth]{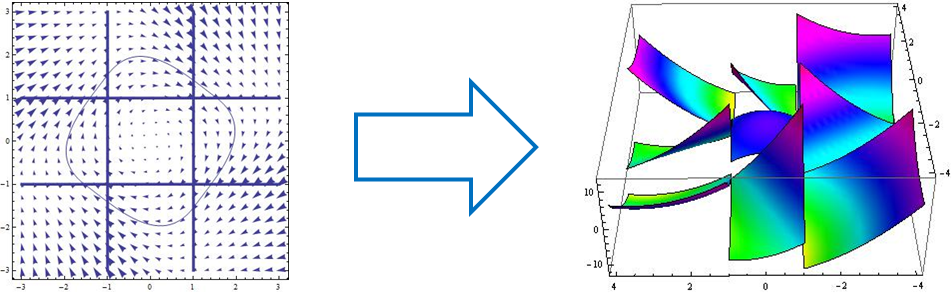}
\end{center}
\caption{
Fig. 1. Multiple QLF approach:
Construct a Lyapunov function for each linear region.
Lyapunov functions in different regions are not mutually comparable.
}
\label{fig:former}
\end{figure}

The multiple QLF aims to focus on the local behaviors of a system in each linear region (see Fig. 1).
This approach constructs a Lyapunov function for each of the linear regions, which is monotonically decreasing along the system's dynamics in the prescribed area.
Note that the functions constructed in the different regions do not equal to each other over the boundaries, and hence lacks global properties.
Thus, the analytical power of multiple QLF is confined only to the local area, instead of being expandable for the dynamics of the entire system.

\begin{figure}
\begin{center}
  \includegraphics[width=0.9\textwidth]{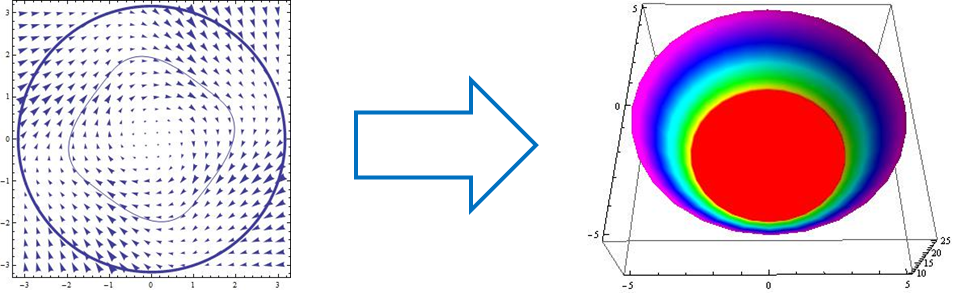}
\end{center}
\caption{
Fig. 2. LaSalle Lyapunov function approach:
Set a LaSalle invariant set and construct Lyapunov function outside the invariant set.
Detailed behaviors inside the invariant set are not considered.
}
\label{fig:former}
\end{figure}

On the contrary, the search for the LaSalle invariant set does focus on the global behaviors of a system.
This effort constructs a Lyapunov function outside of the ``LaSalle invariant set" \cite{LaSalle} (see Fig. 2).
The approach therefore only evaluates global stability of the system, while neglects the detailed behaviors within the invariant set.

\begin{figure}
\begin{center}
  \includegraphics[width=0.9\textwidth]{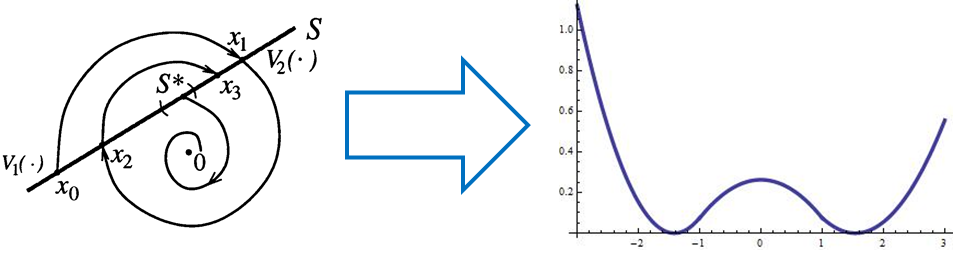}
\end{center}
\caption{
Fig. 3. Surface QLF approach:
Represent the original system by its impact map over the boundaries of linear regions, and find a Lyapunov function for the impact map.
The approach does not construct stability criteria in the system's phase space, but provides inspiration for the construction.
}
\label{fig:former}
\end{figure}

A more recent effort realizes the difficulty of the QLF approach, and thus does not aim to attain a Lyapunov function in a system's complete phase space.
Rather, it represents the original system by its ``impact map" \cite{SLF} (generalized Poincar\'e map) on the boundary between linear regions.
Quadratic Lyapunov function is constructed for this impact map.
Back into the original phase space, this function can be taken as the Lyapunov function on the region boundary.
Therefore, the approach can provide inspiration for the construction of Lyapunov functions.
However, this work itself does not address the original problem, and hence does not provide a stability criteria in the entire phase space.

All the aforementioned quadratic Lyapunov function methods try to describe and analyze complex dynamics with oscillation.
But each fails to describe the entire system in a part of the phase space.
As a result, the Lyapunov functions lack certain properties, restricting their applications.
Take system identification for example, all the QLF constructed cannot distinguish systems with a limit cycle from some systems with multiple fixed points.
This important drawback of the existing methods motivates us to construct Lyapunov functions in the whole phase space for the analysis and control of oscillating systems.

\section{Lyapunov Function in PLS}
First of all, we formally define the Lyapunov function in PLS:

\begin{definition}[Lyapunov Function \cite{ao04potential,Ruoshi}]
\label{def:cand_Lyapunov}

  Let $\Psi:\mathbb{R}^n\xrightarrow{}\mathbb{R}$ be a continuous function. Then $\Psi$ satisfying the following conditions is called a Lyapunov function for the dynamical system $\dot{\mathbf{x}}=\mathbf{f}(\mathbf{x}):\mathbb{R}^n\xrightarrow{}\mathbb{R}^n$.
  \begin{enumerate}[(a)]

  \item $\dot{\Psi}(\mathbf{x})=\frac{d\Psi}{dt}|_\mathbf{x}\leqslant0$ for all $\mathbf{x}\in \mathbb{R}^n$  if  $\dot{\Psi}(\mathbf{x})$ exists;
  \item $\dot{\Psi}(\mathbf{x^*})=0$ if and only if $\mathbf{x}^*\in\mathcal{O}$, where $\mathcal{O}$ is the limit set of the dynamical system: $\dot{\mathbf{x}}=\mathbf{f}(\mathbf{x})$.
  \end{enumerate}
\end{definition}

In this definition, the limit set is not restricted to a fixed point. It can also be a limit cycle, an invariant torus, or a strange attractor.
The function $\Psi$ is thus a Lyapunov function in its general sense.

Locally, positive Lyapunov function $\Psi$ implies the asymptotic orbital stability as stated in the next theorem.

\begin{theorem}
Suppose the Lyapunov function $\Psi$ exists for a dynamical system: $\dot{\mathbf{x}}=\mathbf{f}(\mathbf{x})$.
Suppose further that the limit set $\mathcal{O}$ consists of a single trajectory: $\bar{\mathbf{x}}(t)$;
or the trajectory $\bar{\mathbf{x}}(t)$ is dense in $\mathcal{O}$.
In some neighborhood $U$ of $\mathcal{O}$, if $\Psi$ satisfies the condition:
\begin{enumerate}[(a)]
  \item  $\Psi(\mathbf{x})>0$, for $\mathbf{x}\in U-\mathcal{O}$,
\end{enumerate}
then $\bar{\mathbf{x}}(t)$ is asymptotically orbitally stable.

\end{theorem}

Globally, the convergence region can be extended by the LaSalle invariance principle \cite{LaSalle} to a bounded simply-connected region: $R = \{\mathbf{x}\ |\ \Psi(\mathbf{x}) < M\}$, satisfying: $\Psi(\mathbf{x})$ is differentiable and $\dot{\Psi}(\mathbf{x}) < 0$ for any $\mathbf{x} \in R \setminus \mathcal{O}$.

\begin{remark}
Notice that definition \ref{def:cand_Lyapunov} of Lyapunov function $\Psi$ has no requirement on the positiveness of $\Psi$.
Therefore, unstable and saddle type limit sets can also be discussed within the current framework.
\end{remark}

\section{A Class of PLS as Model System}
In this paper, we will show the construction of Lyapunov functions for a class of PLS with saturation.
The model being discussed emerges from nature and industry and is described as the following:
\begin{align}
    \dot{\mathbf{x}}=\mathbf{f}(\mathbf{x})=W \cdot \mathbf{Sat}(\mathbf{x})-\mathbf{x}\label{system},
\end{align}
where $\mathbf{Sat}(\mathbf{x})$ denotes a saturation linear function acting on each entry of the vector $\mathbf{x}$
\begin{equation}
    Sat(x_i)=
\left\{
    \begin{array}{l}
     Sign(x_i),\ |x_i|>1\\
     x_i,\ |x_i|\leqslant 1.
     \end{array}
\right.\nonumber
\end{equation}

This model quite generally describes the saturating feedback and exponential decay of many natural and technological systems.
Matrix $W$ in equation (\ref{system}) determines the strength of the feedbacks and the structural properties of the network.
When taking time-reversal, the system would be globally unstable. Thus, attracting region problems will manifest \cite{Hu04stabilityregions}.
Moreover, it can be viewed as the linear differential inclusion (LDI) approximation of nonlinear systems with saturation.

To present our results, we choose to be insightful rather than exhaustive.
Hence, we start with two dimensional cases that are $\pi/2$ rotational symmetric in phase space.
Such setting allows clear and simple presentation for the analysis of systems with limit cycle oscillations.
Under this setting, $W$ can be written as $S+T$, where $S=w_{11}\cdot I$ and $T=w_{12} \cdot
\left(
      \begin{array}{cc}
        0& 1  \\
        -1& 0
      \end{array}
\right)$.
The $w_{11}$ and $w_{12}$ defined above represent two degrees of freedom that attracts our interest.
These two degrees of freedom characterize symmetric and antisymmetric feedbacks, which are the two determinant components in oscillating systems.
As will be discussed in section (5), these two components each corresponds to a kind of bifurcation leading to the emergence of limit cycle bifurcation.


The model system has the following different behaviors:
\begin{enumerate}[(a)]
\item
When $w_{11}<1$, there is a global stable fixed point in the phase space;

\item
When $w_{11} \geqslant 1$ and $|w_{12}| \leqslant w_{11}-1$, there are multiple stable fixed points, saddle points and an unstable fixed point;

\item
When $w_{11}>1$ and $|w_{12}|>w_{11}-1$, the system has the behavior of limit cycle.
\end{enumerate}

Among previous works, piecewise QLF have been successfully applied in case (a) and (b) and provided stability measure for the system.
Therefore, we focus on constructing Lyapunov functions in case (c) with limit cycle oscillation in the next section.

\section{Constructing Lyapunov Functions for Oscillating PLS}
We start to construct Lyapunov functions for the PLS with limit cycle when $w_{11}>1$ and $|w_{12}|>w_{11}-1$, addressing a central problem posed by previous efforts mentioned in section (1):
To make Lyapunov function totally continuous in the system's phase space, i.e., Lyapunov function in the neighboring linear regions should equal to each other on the region boundary.

For a system with only fixed points, this continuity problem can be treated as a part of convex optimization problem by the piecewise quadratic Lyapunov function approach.
But for the system with limit cycle, this problem is essentially a periodic boundary condition problem.
Instead of solving a set of partial differential inequalities, we observe that the behavior near the limit set dominates the system's total behavior.
Therefore, the Lyapunov function is constructed in the following three steps:
\begin{enumerate}[(a)]

\item
First, set Lyapunov function equal along the limit cycle of the system to meet the boundary condition in the neighborhood of the limit set.

\item
Second, use reparameterization to deform the Lyapunov function in the linear regions where limit cycle pass through.
After this step, the boundary condition is satisfied between regions containing the limit set.

\item
Third, obtain a totally continuous Lyapunov function by ``gluing" its expressions in all the linear regions with different limit sets together.
\end{enumerate}

\begin{figure}
\begin{center}
  \subfloat[]{\includegraphics[width=0.45\textwidth]{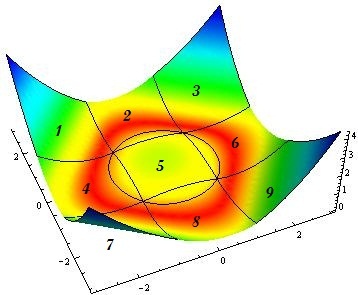}}
  \subfloat[]{\includegraphics[width=0.45\textwidth]{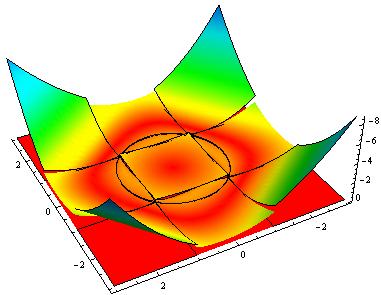}}
\end{center}
\caption{
Fig. 4. Prospective Lyapunov function for the limit cycle system and its Lie derivative:
(a) Prospective Lyapunov function for the limit cycle PLS (constructed in regions numbered from left to right, top to down; labeling from $M_1$ to $M_9$).
(b) Lie derivative of the Lyapunov function (bottom view).
}
\label{fig:construct}
\end{figure}

The resulting Lyapunov function is shown in Panel (a) of Fig. \ref{fig:construct} and the regions are numbered from left to right, top to down; labeling from $M_1$ to $M_9$.

Since the whole system is set as $\pi/2$ rotationally symmetric for convenience, we only need to analyze three regions ($M_2$, $M_3$ and $M_5$ for example) while the other ones are just a change of variables (we can iteratively exchange $(x_2,-x_1)$ for $(x_1,x_2)$ to get the expression of the Lyapunov function for all the other regions).
In the following paragraphs, we take as example the major cases where the limit cycle is contained in the regions: $\{(x_1,x_2),|x_1|>1$ or $|x_2|>1\}$.
The other cases where the limit cycle stays in the regions: $\{(x_1,x_2),|x_1|\leqslant1$ or $|x_2|\leqslant1\}$ can be carried out in the similar way.
And without loss of generality, we can set $w_{12}>(w_{11}-1)$ \footnote{Because in the case of limit cycle, $|w_{12}|>w_{11}-1$.
The case of $w_{12}<1-w_{11}$ can be directly obtained from solving the case: $w_{12}>w_{11}-1$ and changing $(x_2,x_1)$ for $(x_1,x_2)$.}.

In every subsection that follows, we will first explain the method of construction, and then explicitly construct the Lyapunov functions in the linear regions concerned.

\subsection{Step 1}

We first prove that the Lyapunov function should be equal on the limit set (see theorem (\ref{theorem1}) stated below).
This theorem ensures the legitimacy of the first step of construction.

\begin{theorem}
For Lyapunov function $\Psi$ defined on continuous dynamical system with limit sets, $\Psi$ should be equal to a constant on each limit set.\label{theorem1}
\end{theorem}

\begin{remark}
The above theorem states that Lyapunov function reflects the system's stability and convergence towards the limit set.
At the same time, phase information in the neighborhood of the limit set would not be included in the Lyapunov function.
\end{remark}

\begin{figure}
\begin{center}
  {\includegraphics[width=0.9\textwidth]{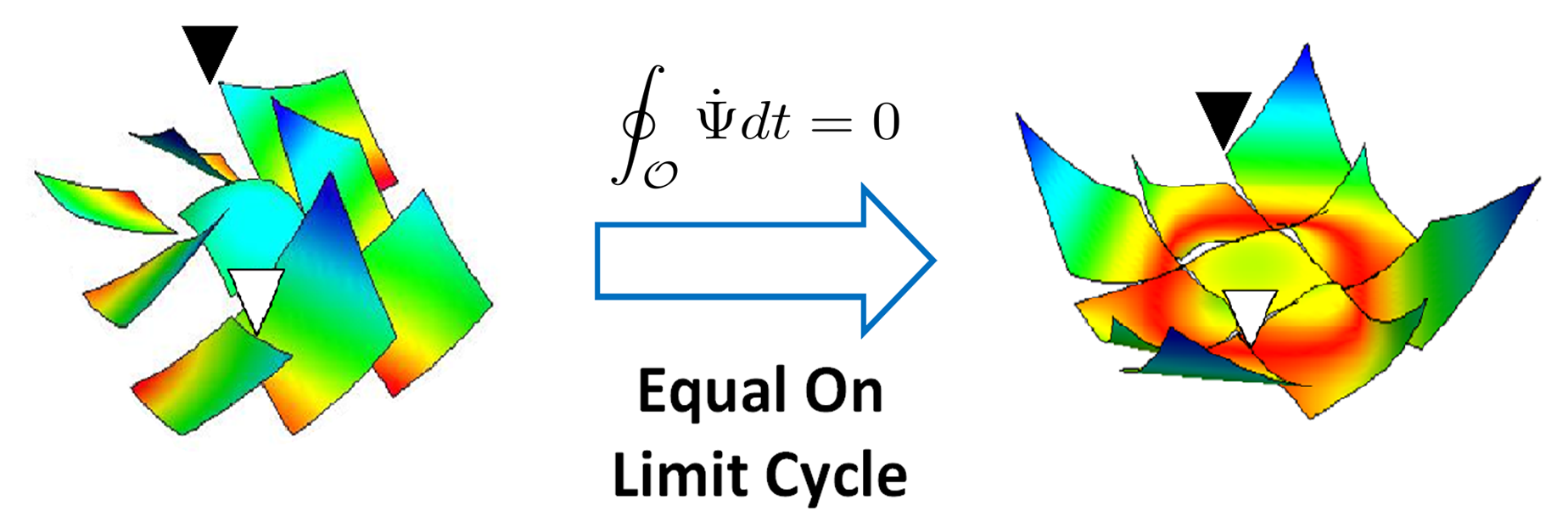}}
\end{center}
\caption{
Fig. 5. Step 1:
We apply the result of theorem $1$ to make Lyapunov function equal on the limit cycle.
White arrow indicates that step $1$ imposes the Lyapunov function to satisfy the boundary condition on the limit cycle, as opposed to multiple QLF method.
Black arrow indicates that the boundary condition is still not met away from the limit cycle.
}
\label{fig:Th1}
\end{figure}

In the neighborhood of the limit cycle, the Lyapunov function is constructed as the follows.

Trajectory of the limit cycle $\mathcal{O}$ can be calculated in each linear region of the system: $M_i$.
With the spatial variables $\mathbf{x}=(x_1,x_2)^\tau$ and proper initial condition $(x_1^0,x_2^0)^\tau$, we denote the trajectory as
\begin{equation}
   \left\{
     \begin{array}{l}
     x_1=F_1(t)\\
     x_2=F_2(t).
     \end{array}
   \right.
\end{equation}

For the systems discussed in this paper, there can be a linear transformation of $x_1$, $x_2$, and accordingly $F_1$, $F_2$:
\begin{equation}
    \left\{
     \begin{array}{l}
     y_1=Y_1(\mathbf{x})=c_{11}x_1+c_{12}x_2
     =c_{11}F_1(t)+c_{12}F_2(t)=G_1(t)\\
     y_2=Y_2(\mathbf{x})=c_{21}x_1+c_{22}x_2
     =c_{21}F_1(t)+c_{22}F_2(t)=G_2(t),
     \end{array}
   \right.\label{G}
\end{equation}
where $G_1$ and $G_2$ can separately have inverse functions with respect to $y_1$ and $y_2$.
Thus, there are two functions $G_1^{-1}(Y_1(\mathbf{x}))$ and $G_2^{-1}(Y_2(\mathbf{x}))$ with the property:
\begin{equation}
     \dfrac{dG_1^{-1}(Y_1(\mathbf{x}))}{dt}=1, \quad \textrm{and} \quad \dfrac{dG_2^{-1}(Y_2(\mathbf{x}))}{dt}=1.
\end{equation}
With this property, the above two functions can represent the phase information of the system in the neighborhood of the limit cycle $\mathcal{O}$.
And note that:
$G_1^{-1}(Y_1(\mathbf{x}))=G_2^{-1}(Y_2(\mathbf{x}))$ if and only if the state
$\mathbf{x}=(x_1,x_2)^\tau \in \mathcal{O}$.
We can construct function $h$ in region $i$ as: $h_i(\mathbf{x})=\exp\left(-G_1^{-1}(Y_1(\mathbf{x}))\right)-\exp\left(-G_2^{-1}(Y_2(\mathbf{x}))\right)$.
Here, $h_i$ is monotonic along the system's dynamics towards the center of the limit cycle $\mathcal{O}$, indicating a series of level curves ``parallel" to $\mathcal{O}$.
Evidently, function $h_i$ reflects the system's convergence towards $\mathcal{O}$ and excludes the phase information along $\mathcal{O}$.

Therefore, we take the Lyapunov function in the neighborhood of the limit cycle $\mathcal{O}$ through region $M_i$ as $\Psi_i$ (with the positive constant $C_i$):
\begin{align}
   \Psi_i=C_i\cdot h_i(\mathbf{x})^2
   =C_i\cdot\left(e^{-G_1^{-1}(Y_1(\mathbf{x}))}-e^{-G_2^{-1}(Y_2(\mathbf{x}))}\right)^2\label{naiveLyapunov}.
\end{align}
It can readily be checked that $\dot\Psi_i(\mathbf{x})=-\Psi_i\leqslant0$, and equality is reached if and only if $\mathbf{x}$ belongs to the limit cycle.
Lyapunov property is thus satisfied in the neighborhood of $\mathcal{O}$.


In the following paragraphs, we explicitly construct Lyapunov functions for the class of the model systems.
As discussed in the previous paragraphs, Lyapunov functions are constructed in each linear region.
In the cases being discussed, region $M_2$, $M_3$, $M_5$ can represent all the regions, and region $M_5$ does not contain any part of the limit cycle.
Hence, we demonstrate the construction in region $M_3$ and $M_2$ in this subsection.


\subsubsection{Region $M_3$}

In region $M_3$ (where $x_1,\ x_2\geqslant1$), the system given in equation (\ref{system}) is:
\begin{equation}
   \left\{
     \begin{array}{l}
     \dot x_1=-x_1+w_{11}+w_{12}\\
     \dot x_2=-x_2+w_{11}-w_{12}.
     \end{array}
   \right.\nonumber
\end{equation}

The segment of limit cycle curve in region $M_3$ is a piece of trajectory starting from a specific initial point $(x_1^0,x_2^0)$.

Corresponding to equation (\ref{G}), we transform $(x_1,x_2)$ into $(y_1,y_2)$:
\begin{equation}
   \left\{
     \begin{array}{l}
     y_1=Y_1(\mathbf{x})=x_1-w_{11}-w_{12}=G_1(t)=y_1^0 \cdot e^{-t}\\
     y_2=Y_2(\mathbf{x})=x_2-w_{11}+w_{12}=G_2(t)=y_2^0 \cdot e^{-t}
     \end{array}
   \right.\nonumber
\end{equation}
and $y_1^0=(x_1^0-w_{11}-w_{12})$; $y_2^0=(x_2^0-w_{11}+w_{12})$.

We can calculate $G_1^{-1}$ and $G_2^{-1}$ as:
\begin{equation}
   \left\{
     \begin{array}{l}
     G_1^{-1}(y_1)=-\log{\dfrac{y_1}{y_1^0}}\\
     G_2^{-1}(y_2)=-\log{\dfrac{y_2}{y_2^0}}.
     \end{array}
   \right.\nonumber
\end{equation}

And $h_3$ would be:
\begin{align}
    h_3(\mathbf{x})
    =e^{-G_1^{-1}(y_1)}-e^{-G_2^{-1}(y_2)}
    =\frac{y_1}{y_1^0}-
    \frac{y_2}{y_2^0}
    =\frac{x_1-w_{11}-w_{12}}{x_1^0-w_{11}-w_{12}}-
    \frac{x_2-w_{11}+w_{12}}{x_2^0-w_{11}+w_{12}}\nonumber.
\end{align}

Consequently, expression of $\Psi_3$ is obtained:
\begin{align}
    \Psi_3=C_3\cdot h_3(\mathbf{x})^2
    =C_3\cdot
    \left(\frac{x_1-w_{11}-w_{12}}{x_1^0-w_{11}-w_{12}}-
    \frac{x_2-w_{11}+w_{12}}{x_2^0-w_{11}+w_{12}}\right)^2.\nonumber
\end{align}


\subsubsection{Region $M_2$}
In region $M_2$ (where $|x_1|<1$ and $x_2\geqslant1$), the system is:
\begin{equation}
   \left\{
     \begin{array}{l}
     \dot x_1=(w_{11}-1)x_1+w_{12}\\
     \dot x_2=-w_{12}x_1-x_2+w_{11}.
     \end{array}
   \right.\nonumber
\end{equation}

We take the similar approach as in region $M_3$. First, $(x_1,x_2)$ is transformed to $(y_1,y_2)$.
And to abbreviate the symbols, we further set $k_1$ as $(w_{11}^2+w_{12}^2)/w_{11}$, and $k_2$ as $w_{12}/(w_{11}-1)$:
\begin{equation}
   \left\{
     \begin{array}{l}
     y_1=Y_1(\mathbf{x})=\left(k_1-\frac{w_{12}}{w_{11}}x_1-x_2\right)\\
     y_2=Y_2(\mathbf{x})=\left(k_2+x_1\right).
     \end{array}
   \right.\nonumber
\end{equation}
Here, $(y_1,y_2)$ is a linearly independent set of variables spanning the state space.
Consequently, the system is transformed to:
\begin{equation}
   \left\{
     \begin{array}{l}
     \dot{y_1}=-y_1\\
     \dot{y_2}=\left(w_{11}-1\right)\cdot y_2.
     \end{array}
   \right.\nonumber
\end{equation}

The limit cycle can be written as:
\begin{equation}
   \left\{
     \begin{array}{l}
     y_1=G_1(t)=y_1^0 \cdot e^{-t}\\
     y_2=G_2(t)=y_2^0 \cdot e^{(w_{11}-1)t}
     \end{array}
   \right.\nonumber
\end{equation}
with $y_1^0=(k_1-(w_{12}/w_{11})x_1^0-x_2^0)$; $y_2^0=(k_2+x_1^0)$.

Just as in region $M_3$, we can have:
\begin{equation}
   \left\{
     \begin{array}{l}
     G_1^{-1}(y_1)=-\log{\dfrac{y_1}{y_1^0}}\\
     G_2^{-1}(y_2)=\dfrac{1}{w_{11}-1}\log{\dfrac{y_2}{y_2^0}}.
     \end{array}
   \right.\nonumber
\end{equation}

And $h_2$ can be expressed as follows:
\begin{align}
    h_2(\mathbf{x})=e^{-G_1^{-1}(y_1)}-e^{-G_2^{-1}(y_2)}
    =\frac{y_1}{y_1^0}-\left(\frac{y_2}{y_2^0}\right)^\frac{1}{1-w_{11}}
    =\frac{k_1-\frac{w_{12}}{w_{11}}x_1-x_2}
    {k_1-\frac{w_{12}}{w_{11}}x_1^0-x_2^0}
    -\left(\frac{k_2+x_1}{k_2+x_1^0}\right)^{\frac{1}{1-w_{11}}}.\nonumber
\end{align}

Now, $\Psi_2$ would be:
\begin{align}
    \Psi_2=C_2\cdot h_2(\mathbf{x})^2
    =C_2\cdot
    \left(\frac{k_1-\frac{w_{12}}{w_{11}}x_1-x_2}
    {k_1-\frac{w_{12}}{w_{11}}x_1^0-x_2^0}
    -\left(\frac{k_2+x_1}{k_2+x_1^0}\right)^{\frac{1}{1-w_{11}}}\right)^2.\nonumber
\end{align}

It's straightforward to check that Lyapunov function $\Psi_i$ constructed in each region is semi-positive definite, with its Lie derivative semi-negative definite and would only equal to $0$ on the limit cycle.

It is also conceivable that $\Psi_i$ is always equal to zero on the limit cycle.
However, away from the limit cycle, $\Psi_i$ and $\Psi_j$ in different regions does not equal to each other on the boundary of $M_i$ and $M_j$.


\subsection{Step 2}

Denote the boundary between region $M_i$ and region $M_j$ as $``\partial M_{i,j}"$.
After the discussion in subsection (1), we can write the boundary condition between two neighboring regions as: $\Psi_i|_{\partial M_{i,j}}=\Psi_j|_{\partial M_{i,j}}$.
It can be observed that the resulting Lyapunov function of step $1$ satisfies: $\Psi_i|_{\partial M_{i,j} \bigcap \mathcal{O}}=\Psi_j|_{\partial M_{i,j}\bigcap \mathcal{O}}=0$.
But, away from the limit cycle $\mathcal{O}$, $\Psi_i|_{\partial M_{i,j}}\neq\Psi_j|_{\partial M_{i,j}}$.
In this subsection, we deform the Lyapunov function $\Psi_i$ and $\Psi_j$ to be equal on the boundary $\partial M_{i,j}$:
$\Psi_i|_{\partial M_{i,j}}=\Psi_j|_{\partial M_{i,j}}$.

To fulfill that aim, we reparameterize the dynamical system.
And the following theorem ensures that reparameterized system can have the same Lyapunov function as the original one.

\begin{theorem}
If two continuous dynamical systems $\dot{\mathbf{x}}=\mathbf{f}(\mathbf{x}), \mathbf{x}\in \mathbb{R}^n$ and
$\dot{\mathbf{x}}=\mathbf{g}(\mathbf{x}), \mathbf{x}\in \mathbb{R}^n$ are orbit equivalent, that is:
\begin{align}
\mathbf{f}(\mathbf{x})=\mu(\mathbf{x})\mathbf{g}(\mathbf{x}),
\end{align}
where $\mu(\mathbf{x})$ is a scalar function and $\mu(\mathbf{x})>0$,
then the Lyapunov function $\Psi$ for one system (if it exists) is also the Lyapunov function for the other system.
\end{theorem}

\begin{remark}
If trajectories $\mathbf{x}(t)$ of the dynamical system: $\dot{\mathbf{x}}=\mathbf{g}(\mathbf{x}), \mathbf{x}\in \mathbb{R}^n$ exists in the whole phase space, then its orbit equivalence with $\dot{\mathbf{x}}=\mathbf{f}(\mathbf{x}), \mathbf{x}\in \mathbb{R}^n$ can also be expressed by an orientation-preserving reparameterization of $``t"$ as $s=s(t)$, where $``s"$ is monotonically increasing and differentiable with respect to $``t"$. (Elaboration of orientation-preserving reparameterization can be found in citation \cite{EDG}.) It can readily be observed that $\mathbf{f}(\mathbf{x})=\frac{d\mathbf{x}}{dt}=\frac{ds}{dt}\cdot\frac{d\mathbf{x}}{ds}
=\mu(\mathbf{x})\frac{d\mathbf{x}}{ds}=\mu(\mathbf{x})\mathbf{g}(\mathbf{x}), \mu(\mathbf{x})>0.$ Thus, one can solve the trajectories of a system, find a proper reparameterization of the parameter $``t"$ and construct Lyapunov function based on it.
\end{remark}

Since $``t"$ in $\mathbf{x}(t)$ is reparameterized to $``s(t)"$, $G_1^{-1}$ and $G_2^{-1}$ are also reparameterized to $s\circ G_1^{-1}$ and $s\circ G_2^{-1}$, so that $h_i(\mathbf{x})$ would be:
\begin{align}
     h_i(\mathbf{x})=\exp\left(-s\circ G_1^{-1}(Y_1(\mathbf{x}))\right)
     -\exp\left(-s\circ G_2^{-1}(Y_2(\mathbf{x}))\right),
\end{align}
where $s$ is a monotonically increasing function.

\begin{figure}
\begin{center}
  {\includegraphics[width=0.9\textwidth]{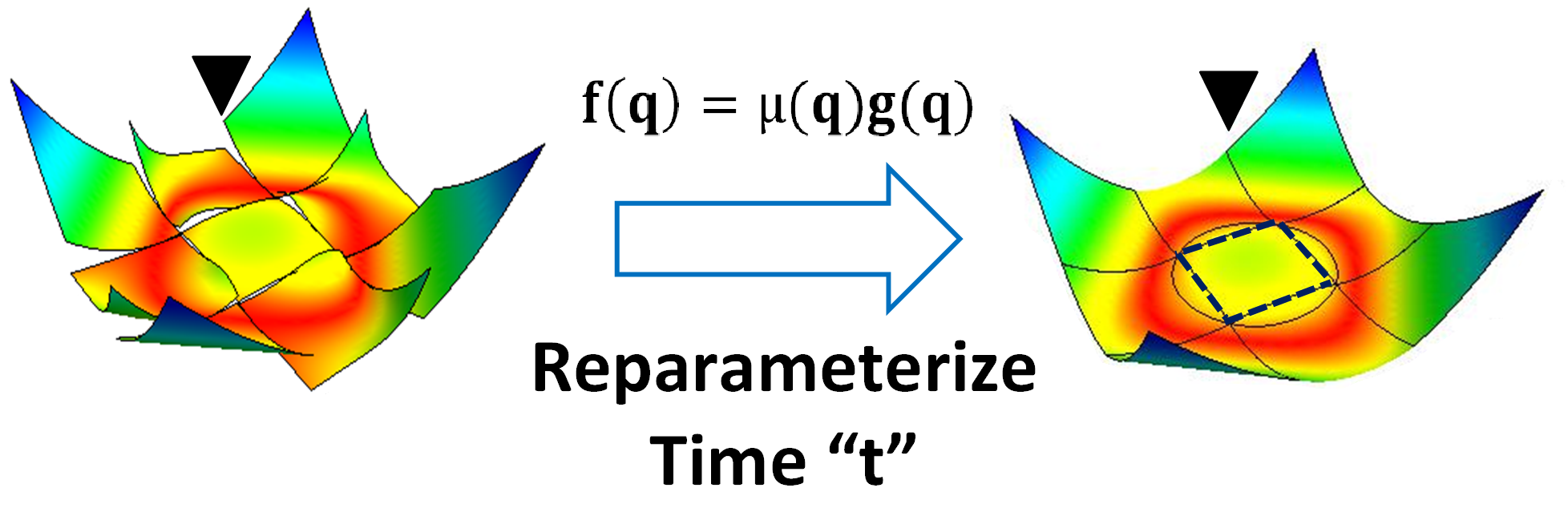}}
\end{center}
\caption{
Fig. 6. Step 2:
In regions containing the limit cycle, we use reparameterization to deform the Lyapunov function.
Black arrow indicates that the function constructed is continuous in regions containing the limit cycle;
Dotted lines indicate that Lyapunov function does not satisfy the boundary condition between the central region (containing a fixed point) and other regions.
}
\label{fig:Th2}
\end{figure}

$\Psi_i$ would thus be reshaped as:
\begin{equation}
   \Psi_i=\frac{1}{2}\left(e^{-s\circ G_1^{-1}(Y_1(\mathbf{x}))}-e^{-s\circ G_2^{-1}(Y_2(\mathbf{x}))}\right)^2,
\end{equation}
such that $\Psi_i|_{\partial M_{i,j}}$ in section $``i"$ would be equal to $\Psi_j|_{\partial M_{i,j}}$ in section $``j"$.


In the following paragraphs, we explicitly perform the second step in the model systems to obtain continuous Lyapunov functions in regions containing the limit cycle.
The resultant Lyapunov function needs to satisfy the condition: $\Psi_1|_{\partial M_{1,2}}=\Psi_2|_{\partial M_{1,2}}$ and $\Psi_2|_{\partial M_{2,3}}=\Psi_3|_{\partial M_{2,3}}$, where $\partial M_{1,2}=\{(x_1,x_2),\ x_1=-1\}$ and $\partial M_{2,3}=\{(x_1,x_2),\ x_1=1\}$.
The ``time" parameter ``$t$" is reparameterized as: ``$s(t)$" in region $M_2$.
In region $M_3$, we simply take $s(t)=t$.

\subsubsection{Region $M_2$}
In region $M_2$, we expect to take $h_2(\mathbf{x})=e^{-s(G_1^{-1})}-e^{-s(G_2^{-1})}$,
and have:
\begin{align}
\Psi_1|_{\partial M_{1,2}}=C_1\cdot h_1(\mathbf{x})^2|_{\partial M_{1,2}}\nonumber
=C_2\cdot h_2(\mathbf{x})^2|_{\partial M_{1,2}}=\Psi_2|_{\partial M_{1,2}};
\end{align}
and
\begin{align}
\Psi_2|_{\partial M_{2,3}}=C_2\cdot h_2(\mathbf{x})^2|_{\partial M_{2,3}}\nonumber
=C_3\cdot h_3(\mathbf{x})^2|_{\partial M_{2,3}}=\Psi_3|_{\partial M_{2,3}},
\end{align}
where $C_2$ and $C_3$ are positive parameters.
Since the reparameterized dynamical system in region three is taken as the same as the original one, $h_3(\mathbf{x})$ here is the same as that in step one, and $h_1(x_1,x_2)=h_3(-x_2,x_1)$.

To obtain the conditions: $\Psi_1|_{\partial M_{1,2}}=\Psi_2|_{\partial M_{1,2}}$, and $
\Psi_2|_{\partial M_{2,3}}=\Psi_3|_{\partial M_{2,3}}$, function $s$ need to have the following initial and final values:
\begin{align}
    s(t)|_{\partial M_{1,2}}=t-\log\left(\lambda_1-\lambda_2\right);\nonumber
\end{align}
and
\begin{align}
    s(t)|_{\partial M_{2,3}}=t-\log\left(\lambda_1+\lambda_2\right),\nonumber
\end{align}
where
\begin{align}
\lambda_1=\frac{k_1-1}{y_1^0}-\frac{1}{2}\left(\frac{k_2-1}{y_2^0}\right)^{\frac{1}{1-w_{11}}}
-\frac{1}{2}\left(\frac{k_2+1}{y_2^0}\right)^{\frac{1}{1-w_{11}}};\nonumber
\end{align}
and
\begin{align}
\lambda_2=\frac{w_{12}/w_{11}}{y_1^0}-\frac{1}{2}\left(\frac{k_2-1}{y_2^0}\right)^{\frac{1}{1-w_{11}}}
+\frac{1}{2}\left(\frac{k_2+1}{y_2^0}\right)^{\frac{1}{1-w_{11}}}.\nonumber
\end{align}

Once the expression of $s(t)$ is obtained, by taking $h_2(\mathbf{x})=e^{-s\circ G_1^{-1}(Y_1(\mathbf{x}))}-e^{-s\circ G_2^{-1}(Y_2(\mathbf{x}))}$ and keeping $h_3(\mathbf{x})$ as it were, we can have
$C_2\cdot h_2(\mathbf{x})^2|_{\partial M_{2,3}}=C_3\cdot h_3(\mathbf{x})^2|_{\partial M_{2,3}}$ as expected.

For the sake of succinctness, we take the function $s(t)$ for reparameterization as:
\begin{align}
    s(t)=t-\log{\left(\lambda_1+\lambda_2 \cdot y_2^0 \cdot e^{(w_{11}-1)t}-k_2\right)}.\nonumber
\end{align}
Here, notations like: $y_1^0$, $y_2^0$, $k_1$, and $k_2$ in this section takes the same meaning as in the previous section: step one.
These are the constants determined by the initial points of the limit cycle in the current region.


%
%
%
%

The spatial expression of $h_2(\mathbf{x})$ can be rewritten as:
\begin{align}
    h_2(\mathbf{x})=\left(\lambda_1+\lambda_2 x_1\right)
    \cdot\left(\frac{k_1-\frac{w_{12}}{w_{11}}x_1-x_2}
    {k_1-\frac{w_{12}}{w_{11}}x_1^0-x_2^0}
    -\left(\frac{k_2+x_1}{k_2+x_1^0}\right)^{\frac{1}{1-w_{11}}}\right).\nonumber
\end{align}

Taking Lyapunov function as the quadratic form $\Psi_2=C_2\cdot h_2(\mathbf{x})^2$ gives:
\begin{align}
    \Psi_2=C_2\cdot\left(\lambda_1+\lambda_2 x_1\right)^2
    \cdot \left(\frac{k_1-\frac{w_{12}}{w_{11}}x_1-x_2}
    {k_1-\frac{w_{12}}{w_{11}}x_1^0-x_2^0}
    -\left(\frac{k_2+x_1}{k_2+x_1^0}\right)^{\frac{1}{1-w_{11}}}\right)^2,
\end{align}
where $C_2=(k_1+w_{12}/w_{11}-x_2^0)^2/(1-w_{11}+w_{12})^2/(\lambda_1+\lambda_2)^2$.

$\Psi_3$ remains unchanged as in the last section (step one): the quadratic form of $h_3(\mathbf{x})$, namely:
\begin{align}
    \Psi_3=h_3(\mathbf{x})^2
    =\left(\frac{x_1-w_{11}-w_{12}}{x_1^0-w_{11}-w_{12}}-
    \frac{x_2-w_{11}+w_{12}}{x_2^0-w_{11}+w_{12}}\right)^2.
\end{align}

It is straightforward to checked that: $\Psi_1$ and $\Psi_2$, $\Psi_2$ and $\Psi_3$ equal to each other on the boundaries.
Lyapunov function is appropriately constructed in the regions containing the limit cycle.

\subsection{Step 3}
The third step of the construction is to fit the Lyapunov function between regions containing different limit sets.
In this case, Lyapunov function in region $M_5$ (containing an unstable fixed point) need to be made equal with that in other regions (containing a stable limit cycle) on the boundaries.
To fulfill this aim, we use the very basic idea of ``gluing technique" \cite{Gluing}, a mathematical approach frequently applied in differential geometry.

The procedure is summarized as follows and explicitly carried out in the model systems later in this section.

\begin{figure}
\begin{center}
  {\includegraphics[width=0.9\textwidth]{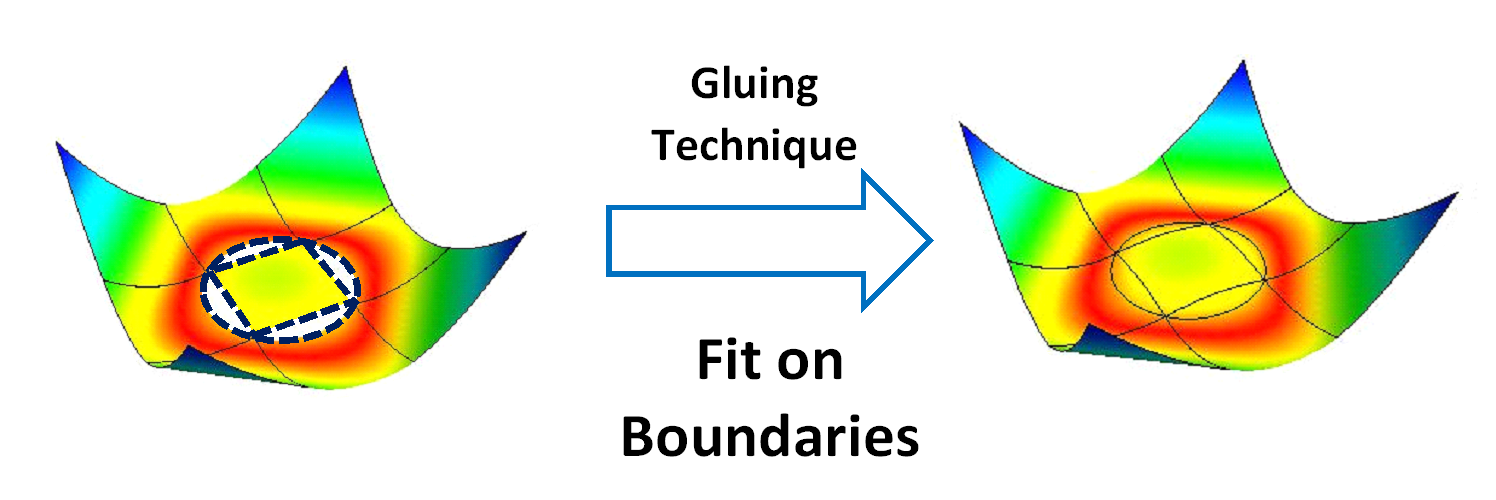}}
\end{center}
\caption{
Fig. 7. Step 3:
First set a gluing region to continuously connect regions on its different sides.
Then integrate along trajectories to solve Lyapunov function for the gluing region.
After step 3, the boundary condition is completely fulfilled.
}
\label{fig:Th3}
\end{figure}

To glue region $M_i$ with $M_j$, first remove a subregion $M_{glue}$ from region $M_i \bigcup M_j$ (as shown in the left part of Fig. \ref{fig:Th3}).
Subregion $M_{glue}$ must contain the boundary $\partial M_{i,j}$ between region $M_i$ and $M_j$: $\partial M_{i,j} \in M_{glue}$.
Also set that $M_{glue}$ does not contain any limit set.
These settings allow Lyapunov function in region $M_i$ to be continuously glued to that in $M_j$ through $M_{glue}$.

Denote the boundary of $M_{glue}$ as $\partial M_{glue}$.
Further denote the part of $\partial M_{glue}$ with vector field flowing into $M_{glue}$ as: $\partial M_{glue}^+$; and the part with vector field flowing out of $M_{glue}$ as: $\partial M_{glue}^-$.
Since $M_{glue}$ does not contain any limit set, trajectories through $M_{glue}$ with starting points $(x_1^0,x_2^0)$ in $\partial M_{glue}^+$ would have end points $(x_1^T,x_2^T)$ in $\partial M_{glue}^-$.

Before finding the expression of the Lyapunov function $\Psi_{glue}$ in $M_{glue}$, it is required that $\Psi|_{(x_1^0,x_2^0)}$ in $\partial M_{glue}^+$ is greater than the corresponding $\Psi|_{(x_1^T,x_2^T)}$ in $\partial M_{glue}^-$.
This can be done by adjusting the free parameters in the expression of Lyapunov function in $M_i$ and $M_j$ as we shall see later in this section.


In the closed region $M_{glue}$,
the expression of $\Psi_{glue}$ can be solved as follows.
Take an arbitrary negative continuous function: $\dot\Psi_{glue}(t,x_1^0,x_2^0)$, expressed with respect to ``time" $t$.
And $\Psi_{glue}$ is:
\begin{align}
    \Psi_{glue}(t,x_1^0,x_2^0)\label{glue}
    =\int_0^t{\frac{\Psi|_{(x_1^T,x_2^T)}-\Psi|_{(x_1^0,x_2^0)}}{\int_0^T\dot\Psi_{glue}(t,x_1^0,x_2^0)dt}}
    \cdot\dot\Psi_{glue}(\tau,x_1^0,x_2^0)d\tau.
\end{align}
Expression (\ref{glue}) can readily be transformed back to the function of $(x_1,x_2)$:
Once a point $(x_1,x_2)$ in phase space is given, initial point $(x_1^0,x_2^0)$ and the relative ``time" $t$ can all be calculated as its function
by solving the intersecting points of $\partial M_{glue}$ with the trajectory through $(x_1,x_2)$.

%

In the paragraphs below, we apply the third step on the model system to obtain a continuous Lyapunov function.
We first find the expression of the Lyapunov function $\Psi_5$ in region $M_5$ (containing an unstable fixed point).
Then we use the third step to make $\Psi_5$ equal to the Lyapunov function in other regions (containing a stable limit cycle) over the boundaries.
Because of the $\pi/2$ symmetry of the system, we only consider boundaries between region $M_5$ and region $M_2$, $M_3$.
And since region $M_3$ is not adjacent to region $M_5$, only boundary $\partial M_{2,5}$ between region $M_2$ and $M_5$ need to be considered.


\subsubsection{Region $M_5$}
In region $M_5$, where $|x_1|,\ |x_2|\leqslant1$,
we construct Lyapunov function as the following to account for the system's dynamics near the central unstable fixed point:
\begin{align}
\Psi_5=C_5\cdot\left(2d+1-d(x_1^2+x_2^2)\right),\label{EQEQ}
\end{align}
where $C_5=(\lambda_1-\lambda_2)^2\cdot(k_1+w_{12}/w_{11}-x_2^0)^2/(1-w_{11}+w_{12})^2$ and $d$ is a free parameter that can be adjusted.

Taking Lie derivative of $\Psi_5$ in $M_5$, one would easily find that
\begin{align}
\dot\Psi_5=-2dC_5\cdot(w_{11}-1)\cdot\left(x_1^2+x_2^2\right),\nonumber
\end{align}
which is less than or equal to $0$, and equality is taken only at the fixed point.

Next, we glue the expressions of Lyapunov function in $M_5$ with that in $M_2$.

\subsubsection{Set $M_{glue}$}
First, a subregion $M_{glue}$ need to be set from $M_2 \bigcup M_5$ for gluing (called gluing region).
Define the region by a closed curve $\partial M_{glue}$ as its boundary.
For convenience, we take the boundary between $M_2$ and $M_5$: $\partial M_{2,5}=\left\{(x_1,x_2),|x_1|\leqslant1,x_2=1\right\}$ as a part of $\partial M_{glue}$.
And we assign the other part of the boundary belonging to region $M_2$: $\partial M_{glue}-\partial M_{2,5} \in M_2$.
Hence, the gluing region is a subregion of region $M_2 \bigcup \partial M_{2,5}$, with no affect on region $M_5$.

Next, we use the result of step 2, the expression of $\Psi_2$, to write the algebraic form of $M_{glue}$.
Since $\Psi_2|_{x_1=1,x_2=1}=\Psi_2|_{x_1=-1,x_2=1}$, the level curve $C$ of $\Psi_2$: $\Psi_2|_{\mathbf{x} \in C}=\Psi_2|_{x_1=-1,x_2=1}$ forms a closed curve with the line segment $\partial M_{2,5}$.
The boundary of region $M_{glue}$ can thus be taken as: $\partial M_{glue} = \partial M_{2,5} \bigcup C$.

On the other hand, since $\nabla\Psi_2\cdot\mathbf{f}_2(\mathbf{x})|_{M_{glue}}=\dot\Psi_2|_{M_{glue}}<0$, vector field on $\partial M_{glue}-\partial M_{2,5}=C$ all flow outward $M_{glue}$.
In other words, $\partial M_{glue}-\partial M_{2,5}\subseteq \partial M_{glue}^-$, and $\partial M_{glue}^+ \subseteq \partial M_{2,5}$.
This setting enables convenient adjustment for the free parameters: we simply take free parameter ``$d$" in the expression of $\Psi_5$ big enough to make $\Psi|_{(x_1^0,x_2^0) \in \partial M_{glue}^+}$ bigger than the corresponding $\Psi|_{(x_1^T,x_2^T) \in \partial M_{glue}^-}$ (where $(x_1^0,x_2^0)$ and $(x_1^T,x_2^T)$ belong to the same trajectory).

So, the gluing region $M_{glue}$ is
\begin{equation}
    \left\{
     \begin{array}{l}
     x_2\geqslant 1\\
     \left(\lambda_1+\lambda_2 x_1\right)^2
     \cdot \left(\frac{k_1-\frac{w_{12}}{w_{11}}x_1-x_2}
     {k_1-\frac{w_{12}}{w_{11}}x_1^0-x_2^0}
     -\left(\frac{k_2+x_1}{k_2+x_1^0}\right)^{\frac{1}{1-w_{11}}}\right)^2
     \leqslant
     \left(\lambda_1+\lambda_2\right)^2
     \cdot\left(\lambda_1-\lambda_2\right)^2
     \end{array}\nonumber
   \right.
\end{equation}
with the equation attained at $\partial M_{glue}$.


Clearly, taking a big parameter $d$ (e.g., $d=w_{11}-1$) in $\Psi_5=C_5\cdot\left(2d+1-d(x_1^2+x_2^2)\right)$ would ensure that $\Psi_5|_{(x_1^0,x_2^0) \in \partial M_{glue}^+} > \Psi_2|_{(x_1^T,x_2^T) \in \partial M_{glue}-\partial M_{2,5}}$.
And since $\Psi_5|_{(x_1^0,x_2^0) \in \partial M_{glue}^+}$ is always bigger than the corresponding $\Psi_5|_{(x_1^T,x_2^T) \in \partial M_{2,5} \bigcap \partial M_{glue}^-}$, we can safely have $\Psi|_{(x_1^0,x_2^0) \in \partial M_{glue}^+} > \Psi|_{(x_1^T,x_2^T) \in \partial M_{glue}^-}$ as requested in this substep.


At last, we ``glue" the expression of the Lyapunov function in region $M_2$ and $M_5$ together by solving $\Psi_{glue}$ in $M_{glue}$.

\subsubsection{Solving $\Psi_{glue}$}

With the trajectories in region $M_{glue}$ given by:
\begin{equation}
\left\{
     \begin{array}{l}
        k_1-\frac{w_{12}}{w_{11}}x_1-x_2=\left(k_1-\frac{w_{12}}{w_{11}}x_1^0-x_2^0\right)\cdot e^{-t}\\
        k_2+x_1=\left(k_2+x_1^0\right)\cdot e^{(w_{11}-1)t}
    \end{array}
\right.,\nonumber
\end{equation}
we
transform $(x_1,x_2)$ into function of $(x_1^0,x_2^0)$ and $t$.
And we take $\dot\Psi_{glue}(t,x_1^0,x_2^0)$ the same expression as $\dot\Psi_2(t,x_1^0,x_2^0)$ (We can also just take $\dot\Psi_{glue}(t)=-1$, but taking $\dot\Psi_{glue}(t)=\dot\Psi_2$ makes the Lyapunov function more smooth on the boundary).
$\Psi_{glue}(t,x_1^0,x_2^0)$ can thus be solved as:
\begin{align}
    \Psi_{glue}(t,x_1^0,x_2^0)
    =\int_0^t{\frac{\Psi|_{(x_1^T,x_2^T)}-\Psi|_{(x_1^0,x_2^0)}}{\int_0^T\dot\Psi_{2}(t,x_1^0,x_2^0)dt}}
    \cdot\dot\Psi_{2}(\tau,x_1^0,x_2^0)d\tau.
\end{align}

We can transform $\Psi_{glue}(t,x_1^0,x_2^0)$ into expression of $\Psi_{glue}(x_1,x_2)$ by solving the intersection of the system's trajectories with $\partial M_{glue}$.

For example, if $(x_1^0,x_2^0)$ is on the $\partial M_{2,5}$ part of the boundary ($x_2=1$), i.e., $(x_1^0,x_2^0)=(x_1^0,1)$; $x_2^0=1$.
$x_1^0=x_1^0(x_1,x_2)$ can be solved inversely by the equation:
\begin{align}
\frac{k_1-\frac{w_{12}}{w_{11}}x_1^0-1}{k_1-\frac{w_{12}}{w_{11}}x_1-x_2}
=\left(\frac{k_2+x_1^0}{k_2+x_1}\right)^{\frac{1}{1-w_{11}}}.\nonumber
\end{align}
Hence,
\begin{align}
t(x_1,x_2)=\frac{1}{w_{11}-1}\log{\frac{k_2+x_1}{x_2+x_1^0(x_1,x_2)}}.\nonumber
\end{align}
Substitute $x_1^0(x_1,x_2)$, $x_2^0(x_1,x_2)$ and $t(x_1,x_2)$ back into $\Psi_{glue}(t,x_1^0,x_2^0)$, we have the expression of $\Psi_{glue}(x_1,x_2)$.

This section gives a detailed guidance on how analytical Lyapunov functions can be constructed in complex PLS with oscillation
(A former effort \cite{Ao06limitcycle} has already shown the possibility of constructing Lyapunov function and relating it with the concept of energy potential in limit cycle systems).
The resulting Lyapunov function and its Lie derivative is shown in Fig. \ref{fig:construct}.
Retrospectively, we can understand that methods derived from quadratic Lyapunov function (QLF) approach cannot be applied in oscillating PLS.
Because Lyapunov functions in systems with limit cycle are not Morse functions, even in a particular region. 

Although in this paper, our methodology is applied in a particular class of PLS for presentation, it's straightforward to see that the approach can be applied to general PLS with limit cycle oscillation.
Moreover, the seemingly calculation intensive procedure not only constructs explicit Lyapunov functions in PLS, but also provides methods for the numerical calculation of Lyapunov functions in other nonlinear dynamical systems.




\section{Emergence of Oscillation from Bifurcations}
With the Lyapunov functions constructed, we can have a geometric view of the PLS.
The geometric configurations of Lyapunov functions not only provides stability measure of the systems \cite{xuwei}, but also describes behavioral changes of the dynamics directly.
This topic has been conceptually discussed in previous works \cite{Zeeman}.
With the explicit expressions of Lyapunov functions, it can be analyzed quantitatively.

From the Lyapunov functions constructed for the class of PLS, we can easily observe two different kinds of bifurcations leading to the emergence of limit cycle oscillation.
One is the change of a stable focus to stable limit cycle, along with the increase of the symmetric feedback: $w_{11}$, called Andronov-Hopf (Hopf) bifurcation \cite{Andronov,Arnold,Marsden}.
The other is the change of multiple stable fixed points to stable limit cycle, along with the increase of the antisymmetric feedback: $w_{12}$, called Saddle-Node-Infinite-Period (SNIP) bifurcation \cite{Shilnikov}.
As the limit cycle emerges from fixed point dynamics, the Lyapunov function changes continuously, indicating the evolution of the system's behavior.

\begin{figure}
\begin{center}
  \includegraphics[width=0.9\textwidth]{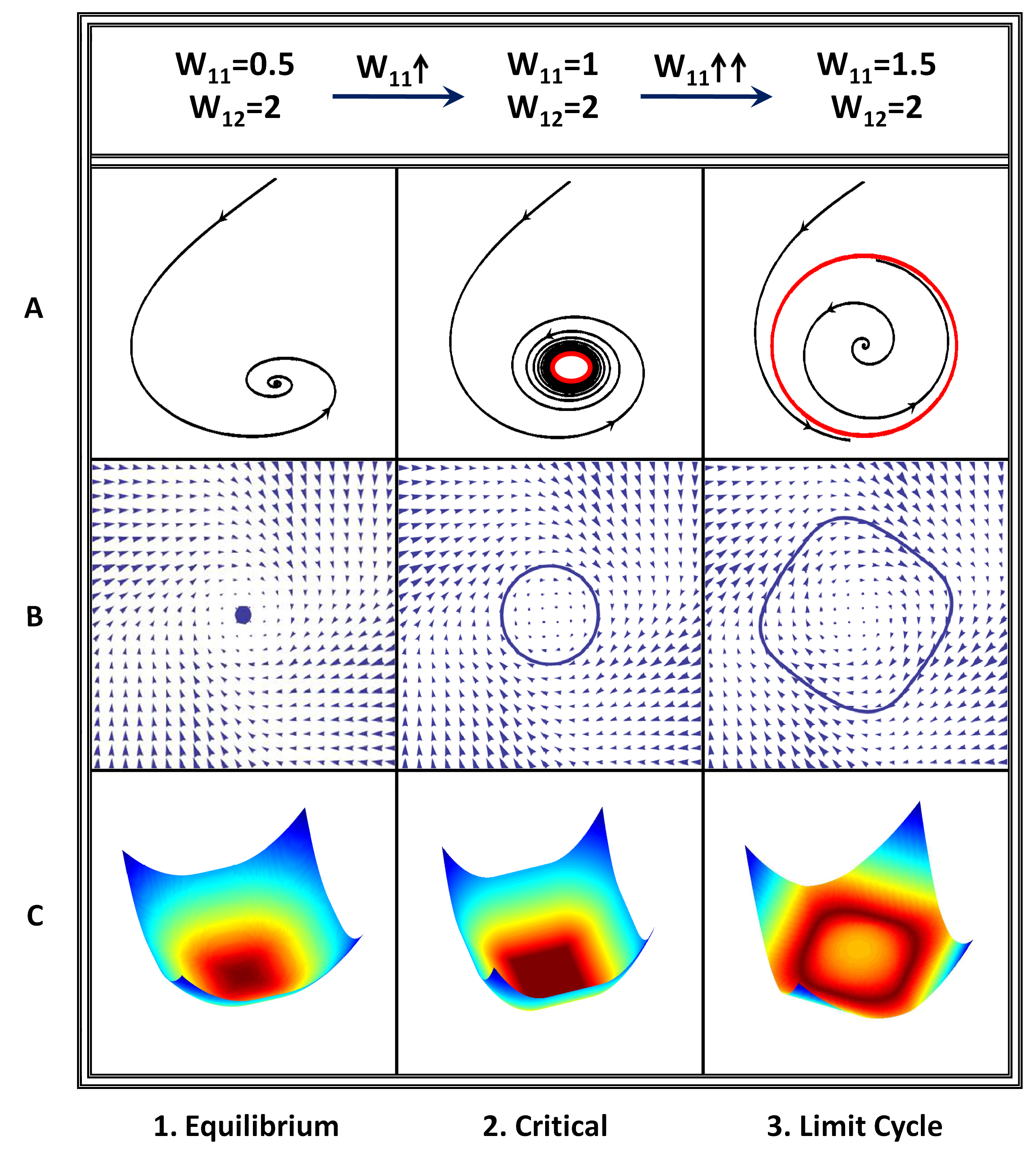}
\end{center}
\caption{
Fig. 8. Hopf bifurcation:
A, Illustrative schemes of Hopf bifurcation: A stable focus changes stability, causing stable limit cycle to form around it.
B, Vector fields of different phases of bifurcation.
C, Lyapunov functions of the according vector fields, indicating how change of symmetric interaction (``s") causes a fixed point to become limit cycle.
}
\label{fig:Hopf}
\end{figure}

\begin{figure}
\begin{center}
  \includegraphics[width=0.9\textwidth]{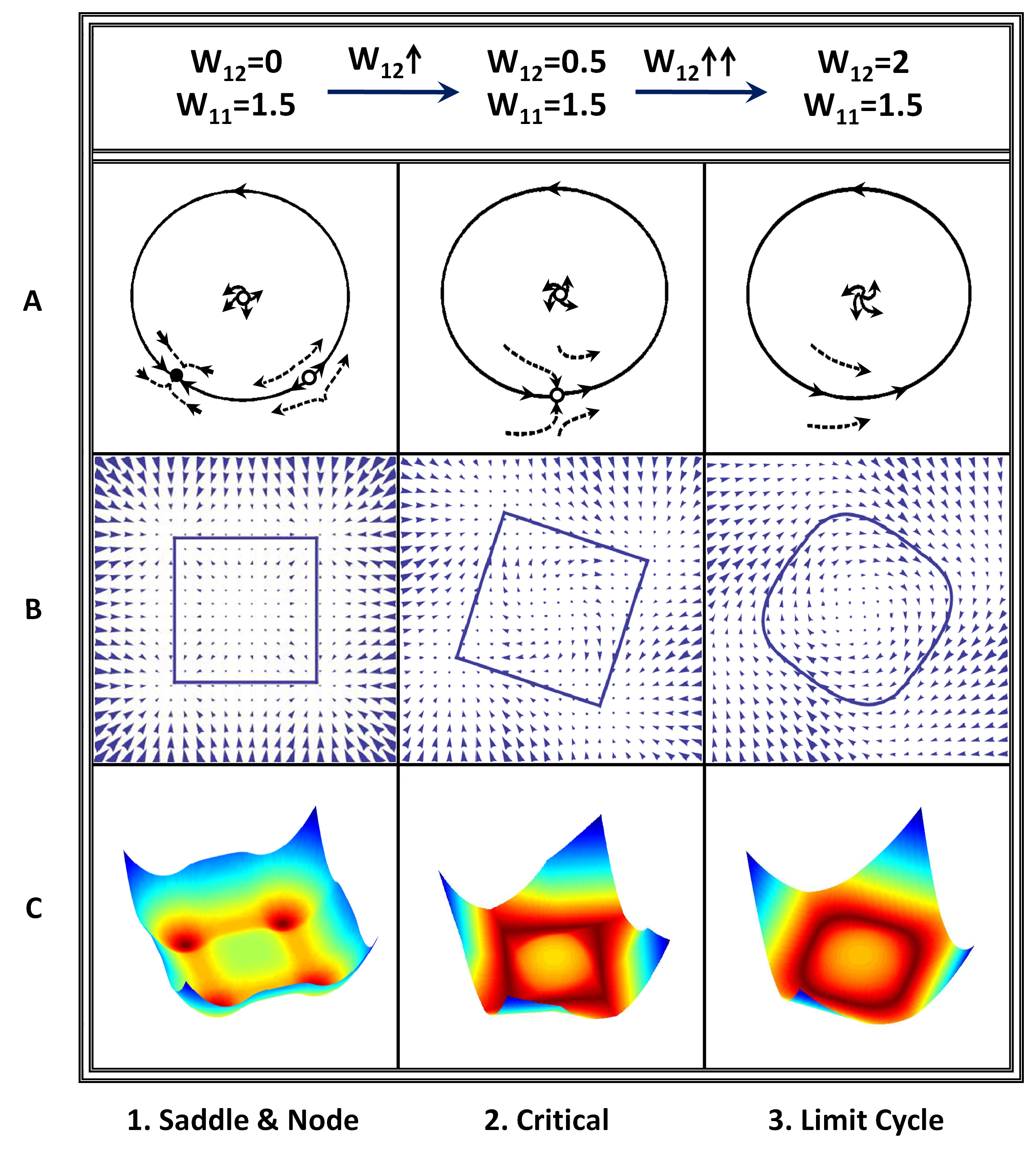}
\end{center}
\caption{
Fig. 9. SNIP bifurcation:
A, Illustrative schemes of SNIP bifurcation: A saddle merge with a node, causing stable limit cycle to emerge.
B, Vector fields of different phases of bifurcation.
C, Lyapunov functions of the according vector fields, indicating how change of antisymmetric interaction (``a") causes a fixed point to become limit cycle.
}
\label{fig:SNIP}
\end{figure}

As symmetric feedback $w_{11}$ of the model system increases, passing the value of $w_{11}=1$ (under the condition of $w_{12}>w_{11}-1$), Hopf bifurcation would happen.
At first, when $w_{11}<1$, the Lyapunov function is a totally convex upward, bowl-shaped function.
This shape indicates that all the states in phase space are attracted to the center.
Then, with the increase of $w_{11}$, Lyapunov function in the center would rise up.
When $w_{11}=1$, at the critical value, Lyapunov function in the central region $M_5$ would have constant value.
Predictably, constant Lyapunov function reflects the conserved dynamics, corresponding to the fact that region $M_5$ is filled with infinite periodic orbits.
The whole region: $x_1^2+x_2^2=1$ is the limit set.
Asymptotic stability of the fixed point is lost.
As $w_{11}$ continues to increase, the Lyapunov function in the center would become concave, rendering the whole function a Mexican-hat shape.
Thus, all the states would converge downward along the Lyapunov function to the limit cycle.

On the other hand, as antisymmetric feedback $w_{12}$ of the model system increases, passing the value of $w_{12}=w_{11}-1$ (under the condition of $w_{11}>1$), SNIP bifurcation would happen.
At first, when $w_{12}<w_{11}-1$, the Lyapunov function has multiple local minima.
Between the local minima, there are saddle points in the Lyapunov function, forming barriers separating the phase space into multiple attracting regions.
With the increase of $w_{12}$
, barriers between attracting regions are lowered.
When $w_{12}=w_{11}-1$, at the critical value, the barriers decrease to zero, all the attracting regions are connected to one.
Four heteroclinic orbits connect to form the set: $\{\mathbf{x} \, | \; \nabla\Psi(\mathbf{x})=0\}$.
And stability of the fixed points is lost.
As $w_{12}$ continues to increase, the limit cycle would be smoothed, allowing the speed of circulation on the limit cycle to increase.

From Fig. \ref{fig:Hopf} and Fig. \ref{fig:SNIP}, we can observe the evolution of the Lyapunov function corresponding to the bifurcation schemes and the changing vector fields.
The continuous change of the Lyapunov function during the bifurcation figuratively explains for the different mechanisms of Hopf and SNIP bifurcation phenomena.
The Hopf bifurcation is essentially the change of stability of the central fixed point, a result of the symmetric feedback exceeding the exponential decay of the system.
While SNIP bifurcation is caused by the increase of rotation effect in phase space, which links different attracting regions together.

\section{Conclusion}
Motivated by the need of Lyapunov functions for complex PLS with limit cycle oscillation, this paper provides a first example of constructive methodology in a class of PLS.


The approach constructs Lyapunov functions in a class of piecewise linear models, tackling a central obstacle faced by previous efforts: to make Lyapunov function of different linear regions continuous at the boundaries.
The construction is completed within three steps:
First, set the Lyapunov function equal on the limit cycle;
second, use reparameterization to obtain continuous Lypuanov function in linear regions containing the limit cycle;
third, glue all the linear regions together to have a totally continuous Lyapunov function in phase space.
The Lyapunov function constructed in this way provides stability measure for the system in its entire phase space.



Moreover, the Lyapunov functions provide a novel geometric point of view on control systems, containing stability measure and behavioral description of the systems' dynamics.
Consequently, the change of Lyapunov functions describes the evolution of the systems' behaviors, explaining for the different factors causing Hopf and SNIP bifurcations respectively.

Clearly, this effort can be extended to the general two dimensional dynamical systems with limit cycle oscillation.
How the approach can be applied in higher dimensional systems with more complex behaviors is still an open problem with great theoretical and practical interest.
Recently, one of our works has shown the possibility of its application in chaotic systems \cite{yian}.

\section*{Appendix\\Proofs}
\begin{proof}[Proof of Theorem 1]
Denote the positive trajectory of $\bar{\mathbf{x}}(t)$ using its initial point: $O^+(x_0,t_0)$.
Suppose $O^+(x_0,t_0)$ is the limit set $\mathcal{O}$; or $O^+(x_0,t_0)$ is dense in $\mathcal{O}$.
%
Then consider a neiborhood of $\mathcal{O}$:
\begin{align}
B_\epsilon(\mathcal{O})=\{\mathbf{x}\in \mathbb{R}^n | d(\mathbf{x} , \mathcal{O}) < \epsilon\}, \nonumber
\end{align}
where $\epsilon$ is chosen so small that $B_\epsilon(\mathcal{O}) \subset U$.
Let $m$ be the minimum value of $\Psi$ on the boundary of $B_\epsilon(\mathcal{O})$.
By premise (a), $m>0$.

Let:
\begin{align}
U_1=\{\mathbf{x}\in B_\epsilon(\mathcal{O}) | \Psi(\mathbf{x})< m\}. \nonumber
\end{align}
Denote $\delta = d(\partial U_1 , \mathcal{O}) = \min_{\mathbf{y}\in \partial U_1} d(\mathbf{y},\mathcal{O})$.
Since $m>0$, by (b) of definition \ref{def:cand_Lyapunov}, $\delta>0$.

Take $\mathbf{y}(t_0) \in B_\delta(\mathcal{O}) \subset U_1$, $\Psi(\mathbf{y}(t_0))< m$.
by (a) of definition \ref{def:cand_Lyapunov}, $\Psi(\mathbf{y}(t))< m$ for any $t > t_0$.
Hence by our construction the trajectory $\mathbf{y}(t)$ cannot leave $U_1 \subset  B_\epsilon(\mathcal{O})$.

Therefore, for any $\mathbf{y}(t)$ satisfying: $d(\mathbf{y}(t_0),\mathcal{O})<\delta$, $d(\mathbf{y}(t),\mathcal{O})<\epsilon$, for any $t>t_0$.

Then, incurring the LaSalle invariance principle, we have: $\lim_{t \rightarrow \infty} d(\mathbf{y}(t) , \mathcal{O}) = 0$.

Since $O^+(x_0,t_0)$ is the limit set $\mathcal{O}$, or $O^+(x_0,t_0)$ is dense in $\mathcal{O}$,
$d(\mathbf{y}(t),O^+(x_0,t_0)) \leqslant d(\mathbf{y}(t),\mathcal{O}) + d(\mathcal{O},O^+(x_0,t_0)) < \epsilon_1$, for any $t>t_0$.
This proves the asymptotic orbital stability of $\bar{\mathbf{x}}(t)$.

\end{proof}

\begin{proof}[Proof of Theorem 2]
Proof of this theorem follows directly from item (b) of the definition of Lyapunov function.
\end{proof}

\begin{proof}[Proof of Theorem 3]
Suppose $\Psi$ is a Lyapunov function for the system $\dot{\mathbf{x}}=\mathbf{g}(\mathbf{x}), \mathbf{x}\in \mathbb{R}^n$, then the following two conditions apply.
\begin{enumerate}[(a)]
\item
If the limit set is simply a fixed point $\mathbf{x}^*$, then $\dot{\mathbf{x}}^*=\mathbf{g}(\mathbf{x}^*)=0$, and $\nabla \Psi(\mathbf{x}^*)=0$. And since $\mu(\mathbf{x}^*)>0$, $\mathbf{f}(\mathbf{x}^*)=\mu(\mathbf{x}^*)\mathbf{g}(\mathbf{x}^*)=0$ if and only if $\mathbf{g}(\mathbf{x}^*)=0$. So, $\nabla \Psi(\mathbf{x}^*)=0$ for $\mathbf{x}^*$ where $\mathbf{f}(\mathbf{x}^*)=0$.

If the limit set is composed of a piece of trajectory: $\mathbf{x}(t)$, then there exists an orientation-preserving reparameterization of $t$: $s=s(t)$, such that $\mathbf{f}(\mathbf{x})=\dfrac{d\mathbf{x}}{dt}=\dfrac{ds}{dt}\cdot\dfrac{d\mathbf{x}}{ds}
=\mu(\mathbf{x})\dfrac{d\mathbf{x}}{ds}=\mu(\mathbf{x})\mathbf{g}(\mathbf{x})$, where $s$ is monotonically increasing and differentiable almost everywhere with respect to $t$.
Hence, $\mathbf{x}^*\in\mathcal{O}$, where $\mathcal{O}$ is a limit set for the system $\dot{\mathbf{x}}=\mathbf{f}(\mathbf{x})$ if and only if $\mathcal{O}$ is also a limit set for the system $\dot{\mathbf{x}}=\mathbf{g}(\mathbf{x})$. So, $\nabla \Psi(\mathbf{x^*})=0$ if and only if $\mathbf{x}^*\in\mathcal{O}$ where $\mathcal{O}$ is the limit set of the dynamical system: $\dot{\mathbf{x}}=\mathbf{f}(\mathbf{x})$.

Until here, item (a) of the definition of Lyapunov function is proved.
\item
$\dot{\Psi}(\mathbf{x})=\frac{d\Psi}{dt}|_\mathbf{x}=d\Psi(\mathbf{g}_{\mathbf{x}})/dt\leqslant0$ for all $\mathbf{x}\in \mathbb{R}^n$  if  $\dot{\Psi}(\mathbf{x})$ exists. So, for the system: $\dot{\mathbf{x}}=f(\mathbf{x}), \mathbf{x}\in \mathbb{R}^n$, $\dot{\Psi}(\mathbf{x})=d\Psi(\mathbf{f}_{\mathbf{x}})/dt=\mu(\mathbf{x})d\Psi(\mathbf{g}_{\mathbf{x}})/dt\leqslant0$.

Up to here, item (b) of the definition of Lyapunov function is proved.
\end{enumerate}
Thus, $\Psi$ is also a Lyapunov function for the system $\dot{\mathbf{x}}=\mathbf{f}(\mathbf{x}), \mathbf{x}\in \mathbb{R}^n$.
\end{proof}

\section*{Acknowledgement}
The authors would like to express their sincere gratitude
to Xinan Wang, Ying Tang, Tianqi Chen, Jianghong Shi and Song
Xu for their constructive advice.


\end{document}